\documentclass[aps,prb,showpacs,twocolumn,superscriptaddress]{revtex4-1}

\usepackage{amssymb}
\usepackage{graphicx}
\usepackage{amsmath}
\usepackage{color}
\usepackage{hyperref}

\def\hfqmc{\mbox{HF-QMC} }
\def\s{\sigma}

\def\c{\hat{c}^{}}
\def\cc{\hat{c}^{+}}

\begin{document}

\title{Magnetism of iron and nickel from rotationally invariant \\ Hirsch-Fye quantum Monte Carlo calculations}

\author{A. S. Belozerov}
\affiliation{Institute of Metal Physics, Russian Academy of Sciences, 620990 Yekaterinburg, Russia}
\affiliation{Ural Federal University, 620990 Yekaterinburg, Russia}

\author{I. Leonov}
\affiliation{Theoretical Physics III, Center for Electronic Correlations and Magnetism, Institute of Physics, University of Augsburg, 86135 Augsburg, Germany}

\author{V. I. Anisimov}
\affiliation{Institute of Metal Physics, Russian Academy of Sciences, 620990 Yekaterinburg, Russia}
\affiliation{Ural Federal University, 620990 Yekaterinburg, Russia}

\date{\today}

\begin{abstract}

We present a rotationally invariant Hirsch-Fye quantum Monte Carlo algorithm in which
the spin rotational invariance of Hund's exchange is approximated by averaging
over all possible directions of the spin quantization axis.
We employ this technique to perform benchmark calculations for the two- and three-band Hubbard models on the
infinite-dimensional Bethe lattice. Our results agree quantitatively well with those obtained using
the continuous-time quantum Monte Carlo method with rotationally invariant Coulomb interaction. 
The proposed approach is employed to compute the electronic
and magnetic properties of paramagnetic $\alpha$ iron and nickel. The obtained Curie temperatures agree well
with experiment. Our results indicate that the magnetic transition temperature is significantly overestimated
by using the density-density type of Coulomb interaction.

\end{abstract}

\maketitle

\section{Introduction}

The theoretical description of the electronic properties of transition metal 
compounds with partially filled $d$ and $f$-shells and strong Coulomb interaction 
between the electrons remains a challenging, fundamental problem in condensed matter 
physics.~\cite{DMFT,Correlated_compounds} The interplay between electronic and lattice 
degrees of freedom in such materials results in their diverse physical properties
and rich phase diagrams making these compounds particularly attractive for technological 
applications.~\cite{Imada98} Moreover, orbital degeneracy is an important and often 
inevitable cause of this complexity. Together with the Hund's exchange interaction,
it has important implications for the electronic and magnetic properties of correlated 
materials, leading to formation of local moments and complicated multiplet structures.

The electronic properties of correlated materials can be understood
by employing the so-called LDA+DMFT approach,~\cite{DMFT,LDA+DMFT}
a combination of \textit{ab initio} local density approximation (LDA) 
of the density functional theory and dynamical mean-field theory (DMFT).
Nowadays, the LDA+DMFT technique has become a state-of-the-art method
for realistic description of correlated electron materials from first 
principles. This approach provides a systematic many-body treatment of 
the effect of local electronic correlations by taking into account temporal 
fluctuations while spatial fluctuations are neglected. Applications of 
LDA+DMFT for correlated electron compounds such as transition metals and 
their oxides have provided important insights into our understanding of 
the electronic and magnetic properties of these materials. In particular, 
by employing the LDA+DMFT technique it has become possible to obtain a 
good quantitative description of localized as well as delocalized electron 
states. In addition, the approach allows one to determine the electronic and 
magnetic properties of correlated compounds in both paramagnetic and magnetically 
ordered states.

Nevertheless, there are two important limitations of conventional implementations of 
the LDA+DMFT method.
The first originates from the single-site (local) nature of DMFT.
In particular, the key assumption of the theory is the limit of infinite spatial dimension,
which allows one to perform an exact mapping of a complex lattice model (such as the Hubbard 
model) to a quantum impurity with an energy-dependent external bath, resulting in $k$-independent 
self-energy. However, in some cases the non-local spatial correlations can be essential
to provide a correct description of the properties of correlated materials.~\cite{Extensions_applic}
For instance, the standard LDA+DMFT calculations are not able to capture the reduction of magnetic 
transition temperature due to long-wavelength spin waves.
To resolve this problem several methods have been recently proposed,~\cite{DMFT_extensions}
which we leave beyond the scope of our paper.

The second limitation concerns the spin rotational symmetry
of the Hund's exchange interaction.
Since correlated materials often have several bands at the Fermi level,
it requires specific treatment of the local Coulomb interaction,
which in a cubic environment consists of the intra- and inter-orbital
Coulomb interactions $U$ and $U'$, Hund's exchange $J$, and the pair-hopping 
coupling $J'$. These interactions obey spin and orbital rotational symmetry,
thereby ${U=U'+J+J'}$ ensures the rotational invariance in the orbital space
and ${J=J'}$ can be assumed whenever the spin-orbital coupling is negligible.
Unfortunately, it is difficult to handle all these multiband interactions
including Hund's exchange coupling and the pair hopping term with the quantum 
Monte Carlo (QMC) method. In particular, a straightforward implementation leads 
to a severe sign problem making such simulations unfeasible. Therefore, at present, 
the most material-specific calculations employ the approximate form of the Coulomb 
repulsion restricted to the Ising-type exchange interaction. These calculations 
often provide a good quantitative description of the electronic, magnetic, and 
structural properties of correlated materials
as a function of the reduced temperature ${T/T_C}$, where $T_C$
is the calculated temperature of magnetic ordering.~\cite{Licht01,Fe_Leonov}
However, the correct symmetry of the exchange interaction turns out to be essential 
for quantitative description of the electronic and magnetic properties of correlated 
systems.~\cite{Prushke_2band_model,Antipov_3band_model,Sakai,Antipov_2band_model,jqmc,Sakai07}

This problem can be overcome by using some quantum impurity solvers such as numerical 
renormalization group (NRG),~\cite{nrg_solver} exact diagonalization (ED),~\cite{ed_solver}
continuous-time quantum Monte Carlo (CT-QMC),~\cite{CT-QMC} and others,~\cite{Sakai04,Sakai,jqmc}
which allow one to treat the Coulomb interaction in its general form with preserved spin 
rotational symmetry. These calculations performed for the two- and three-band Hubbard models
on the infinite-dimensional Bethe~\cite{Antipov_2band_model,Antipov_3band_model}
and hypercubic~\cite{Sakai} lattices show a substantial overestimation of the magnetic 
transition temperature for the approximate Ising-type form of the exchange Coulomb 
interaction with respect to the rotationally invariant one.
In accordance with this, recent LDA+DMFT calculations of correlated compounds
also indicate that the magnetic transition temperatures appear to be
significantly overestimated by using the density-density type of Coulomb 
interaction.~\cite{Licht01,Kunes_Belozerov,jqmc}
However, applications of these techniques so far have been mostly limited
to simple model systems and only a few realistic calculations for 3$d$ 
compounds have been recently presented.~\cite{Full_U_applic}
This is mostly because of high 
computational costs (exponential with the number of orbitals)
of these methods implemented with the full rotationally invariant
Coulomb interaction which makes such calculations for $3d$ and $4f$ 
materials extremely expensive.
Obviously, the LDA+DMFT investigations of correlated materials with the Coulomb 
interaction in its general form with preserved spin rotational symmetry remain 
problematic and pose a great theoretical challenge.

In this paper, we present an implementation of the LDA+DMFT approach which
allows us to take into account rotational symmetry of the exchange Coulomb interaction.
This approach is formulated in terms of the Hirsch-Fye quantum Monte Carlo algorithm~\cite{hfqmc}
where the spin rotational invariance of Hund's exchange is approximated by averaging over all
possible directions of the spin quantization axis. It provides a robust and computationally 
efficient method which allows us to simulate the five-orbital systems at high temperatures. 
Using this technique we perform benchmark calculations for the two- and three-band Hubbard models on
the infinite-dimensional Bethe lattice. In addition, we employ the proposed approach to 
calculate the electronic and magnetic properties of paramagnetic $\alpha$ iron and nickel.
To outline the importance of rotational symmetry of the exchange Coulomb interaction we 
compare our results with those obtained by using the density-density approximation.

This paper is organized as follows. In Sec.~\ref{sec:method} we present a 
detailed formulation of the proposed approach which allows one to treat rotational 
invariance of the exchange interaction. In Sec.~\ref{sec:results} we employ this 
technique to compute the electronic and magnetic properties of the two- and three-band models 
on the Bethe lattice, paramagnetic $\alpha$ iron, and nickel. The obtained results 
are compared with those of previous calculations and experimental data. 
Finally, conclusions are presented in Sec.~\ref{sec:conclusions}.

\section{Method\label{sec:method}}

The multiband Hamiltonian with full rotationally invariant on-site Coulomb interaction
can be written in the following form:~\cite{Footnote}
\begin{eqnarray} \label{full_ham}
\hat{H}& = & U \sum_{m} \hat n_{m\uparrow}\hat n_{m\downarrow} \\ \nonumber
&+& \dfrac{1}{2} \sum_{\scriptstyle mm\prime\s \atop \scriptstyle m \ne m^\prime} \{ (U-2J) \hat n_{m\s}\hat n_{m^\prime\overline{\s}}
 +  (U-3J) \hat n_{m\s} \hat n_{m^\prime\s} \\ \nonumber
&-& J (\hat c^{\dagger}_{m\s} \hat c^{}_{m\overline{\s}}
    \hat c^{\dagger}_{m\prime\overline{\s}} \hat c^{}_{m\prime\s}
  + \hat c^{\dagger}_{m\s} \hat c^{\dagger}_{m\overline{\s}}
    \hat c^{}_{m\prime\s}  \hat c^{}_{m\prime\overline{\s}}) \},
\end{eqnarray}
where $\cc_{m\s}$ ($\c_{m\s}$) denotes the creation (annihilation) operator
of an electron with spin~$\s$~(${=\uparrow,\downarrow}$) at orbital~$m$,
${\hat n_{m\s} = \cc_{m\s} \c_{m\s}}$,
$U$ is the screened Coulomb interaction parameter,
and $J$ is the Hund's exchange coupling.
The first three terms in Hamiltonian~(\ref{full_ham}) correspond to the density-density part of
Coulomb interaction and contain the exchange interaction in the Ising-type form.
The remaining part consists of spin-flip (4th) and pair hopping (5th) terms.
Using the $z$-projection of the spin operator, ${\hat S_m^z = (\hat n_{m\uparrow}-\hat n_{m\downarrow})/2}$,
and the orbital occupancy operator, ${\hat N_m = \hat n_{m\uparrow}+\hat n_{m\downarrow}}$,
the density-density part can be rewritten as
\begin{eqnarray} \label{dd_ham}
\hat{H}_{\rm{dd}} &=& U \sum_{m} \hat n_{m\uparrow}\hat n_{m\downarrow} \\ \nonumber
&+& \dfrac{1}{2} \sum_{\scriptstyle mm\prime \atop \scriptstyle m \ne m^\prime}
\{\bar U \hat N_m\hat N_{m^\prime}-2J \hat S_{m}^z \hat S_{m^\prime}^z\},
\end{eqnarray}
where $\bar{U}=U-5J/2$\, is the average value of the Coulomb interaction.
The spin-flip term in Eq.~\eqref{full_ham} can be expressed via operators
${\hat S_m^x = (\hat c^\dag_{m\uparrow}\hat c^{}_{m\downarrow} + \hat c^\dag_{m\downarrow}\hat c^{}_{m\uparrow})/2}$
and
${\hat S_m^y = -i (\hat c^\dag_{m\uparrow}\hat c^{}_{m\downarrow} - \hat c^\dag_{m\downarrow}\hat c^{}_{m\uparrow})/2}$
as
\begin{eqnarray}
\sum_\s \hat c^{\dagger}_{m\s} \hat c^{}_{m\overline{\s}}
\hat c^{\dagger}_{m\prime\overline{\s}} \hat c^{}_{m\prime\s}
= 2( \hat S_{m}^x\hat S_{m^\prime}^x + \hat S_{m}^y\hat S_{m^\prime}^y).
\end{eqnarray}
The pair hopping term acts only on high energy states with two electrons on the same 
orbital and thereby can be neglected. However, taking into account the spin-flip term 
in Eq.~(\ref{full_ham}) is crucial for the correct description of spin dynamics.
Therefore, the Coulomb interaction Hamiltonian can be written as
\begin{eqnarray} \label{TO7}
\hat{H} &=& U \sum_{m} \hat n_{m\uparrow}\hat n_{m\downarrow} \\ \nonumber
&+& \dfrac{1}{2} \sum_{\scriptstyle mm\prime \atop \scriptstyle m \ne m^\prime}
\{\bar U \hat N_m\hat N_{m^\prime}-2J \hat{\vec S}_{m} \hat{\vec S}_{m^\prime} \}.
\end{eqnarray}
Hamiltonian (4) with exchange term taken as a vector product,
$J \hat {\vec S}_{m}\hat {\vec S}_{m^\prime }$, is invariant with respect to
the spin quantization axis rotations while the density-density counterpart 
with the Ising-type exchange term, $J \hat S_m^z \hat S_{m^\prime}^z$, is not.

To restore the spin rotational symmetry of Hamiltonian (\ref{dd_ham})
we here employ the 
method originally proposed by Hubbard.~\cite{Hubbard79} The Coulomb interaction in 
Ref.~\onlinecite{Hubbard79} was considered in the following form:
\begin{eqnarray} \label{from_Hubbard}
 U\hat N_{\uparrow}\hat N_{\downarrow}
=\frac{1}{4}U\hat{N}^2-U \hat S_{z}^2
=\frac{1}{4}U\hat{N}^2-U (\vec{e}\hat {\vec S})^2,
\end{eqnarray}
where ${\hat{N}_\s = \sum_m \hat{n}_{m\s}}$,
${\hat{N}=\hat{N}_\uparrow + \hat{N}_\downarrow}$,
${\hat{S}_z=\sum_m \hat{S}_m^z}$,
${{\vec S}}$ is the total spin of the atom, and
$\vec{e}$\; is an arbitrary unit vector that can be interpreted as a quantization axis direction.
In order to restore the spin rotational symmetry in Eq.~(\ref{from_Hubbard}),
averaging over all possible directions of $\vec{e}$ was introduced using
the functional integral technique.~\cite{Hubbard79}
Here we implement this method with the Hirsch-Fye quantum Monte Carlo 
algorithm (\mbox{HF-QMC}).~\cite{hfqmc}

The~\hfqmc is based on the discrete Hubbard-Stratonovich transformation
which employs the identity
\begin{eqnarray} \label{Hub-Strat}
\textrm{exp}[ -\Delta\tau && U_{\mu\nu} \{ \hat{n}_\mu \hat{n}_\nu - \dfrac{1}{2}( \hat{n}_\mu + \hat{n}_\nu ) \} ] \nonumber \\ 
&& = \dfrac{1}{2} \sum_{s_{\mu\nu}=\pm 1} \textrm{exp} \{ \lambda_{\mu\nu} s_{\mu\nu} ( \hat{n}_\mu - \hat{n}_\nu ) \},
\end{eqnarray}
where
$\mu$ and $\nu$
are combined spin-orbital indices,
$s_{\mu\nu}$ is an Ising-like variable taking the values $\pm 1$,
$U_{\mu\nu}$ stands for the matrix element of the Coulomb interaction operator,
and ${\lambda_{\mu\nu} = \textrm{arcosh}[\textrm{exp}({\Delta\tau U_{\mu\nu}/2})]}$.
The imaginary time interval $[0,\beta]$ is divided into $L$ slices of length $\Delta\tau$,
so that $\tau_l=l\Delta\tau$, where $l=1,2,..L$, and $\beta$ denotes the inverse temperature.

Using the Trotter decomposition, the partition function of the system can be approximated as
\begin{eqnarray} \label{PF_init}
 Z = \textrm{Tr}\; e^{-\beta(\hat{H}_0 + \hat{H}_\textrm{int})} 
   = \textrm{Tr} \prod_{l=1}^L e^{-\Delta\tau(\hat{H}_0 + \hat{H}_\textrm{int})} \nonumber  \\
   \simeq  Z^{\Delta\tau} \equiv \textrm{Tr} \prod_{l=1}^L e^{-\Delta\tau\hat{H}_0} e^{-\Delta\tau\hat{H}_\textrm{int}}, \qquad
\end{eqnarray}
where $\hat{H}_0$ is the non-interacting (quadratic in fermion operators) part of Hamiltonian (1) and $\hat{H}_\textrm{int}$ 
describes the Coulomb interaction.
%
%
Therefore, the partition function can be written as a sum over all auxiliary field configurations:
\begin{eqnarray} \label{PF}
  Z^{\Delta\tau} = \dfrac{1}{2^{N_fL}} \sum_{\{s\}=\pm 1} z(s),
\end{eqnarray}
where $z(s)$ is the partition function for a particular configuration of auxiliary fields,
$\{s\}$ denotes the set of all auxiliary fields,
and ${N_f=M(2M-1)}$ is the number of auxiliary fields for $M$ orbitals.

In the case of the density-density form of the local Coulomb interaction,
the single-electron dynamical potential has only the $z$ component
and can be expressed as
\begin{eqnarray} \label{P1}
V_{\mu} (\tau_l)& = &\sum_{\nu(\ne\mu)}\lambda_{\mu\nu}s_{\mu\nu}(\tau_l) \sigma_{\mu\nu},\\
\sigma_{\mu \nu}&=& \left\{\begin{array}{ll}
 1, &   \, \mu<\nu \\
-1, &    \, \mu>\nu
\end{array} \;. \right.
\end{eqnarray}
For the quantization axis defined by polar angle $\theta$ and azimuthal angle $\phi$,
the single-electron dynamical potential can be written as
\begin{eqnarray} \label{nondiagpot}
V'(\tau) = T^\dag(\theta,\phi)\,V(\tau)\,T(\theta,\phi),
\end{eqnarray}
where $V(\tau)$ is the potential calculated by Eq.~(\ref{P1}) for the quantization axis chosen to coincide with the $z$ axis,
and $T(\theta,\phi)$ stands for a transformation matrix in the spin variables and reads as
\begin{eqnarray} \label{U-theta}
T(\theta,\phi)&=&
\left(\begin{array}{cc}
\cos(\theta/2)e^{i\phi/2}& \sin(\theta/2)e^{-i\phi/2}\\
- \sin(\theta/2)e^{i\phi/2}&\cos(\theta/2)e^{-i\phi/2}
\end{array} \right).
\end{eqnarray}
By integrating over all possible directions of the quantization axis, 
we obtain the partition function of the system with preserved spin rotational 
symmetry:
\begin{eqnarray} \label{new_PF}
  \!\!\!\!\! \widetilde{Z}^{\Delta\tau} = \dfrac{1}{2^{1+N_fL}\,\pi^2}
  \sum_{\{s\}=\pm 1} \int_0^{2\pi} d\phi \int_0^\pi d\theta\; z(s,\theta,\phi).
\end{eqnarray}
Here, $z(s,\theta,\phi)$ is the partition function for a particular configuration of auxiliary fields
with the quantization axis defined by angles $\theta$ and $\phi$.
Similarly to the original HF-QMC algorithm, it can be demonstrated that
\begin{eqnarray} \label{new_PF2}
  z(s,\theta,\phi) = \textrm{det}[G^{-1}(s,\theta,\phi)],
\end{eqnarray}
where $G(s,\theta,\phi)$ is the Green's function for a particular configuration of auxiliary fields,
angles $\theta$ and $\phi$.
The resulting Green's function has the following form
\begin{eqnarray} \label{new_GF}
   \!\!\!\!\!\!\!\! \widetilde{G}^{\Delta\tau} =\dfrac{C}{\widetilde{Z}^{\Delta\tau}}
  \! \sum_{\{s\}=\pm 1} \int_0^{2\pi} \!\!\! d\phi \int_0^\pi \!\! d\theta\, G(s,\theta,\phi)\, z(s,\theta,\phi),
\end{eqnarray}
where ${C=1/(2^{1+N_fL}\,\pi^2)}$.
To integrate over all auxiliary field configurations and all possible directions of the quantization axis
in Eq.~(\ref{new_GF}), we employ the quantum Monte Carlo technique in which\, 
${\textrm{det}[G^{-1}(s,\theta,\phi)]}$ is interpreted as a stochastic weight.
Similarly to the original \hfqmc algorithm, the Green's functions
of two configurations with potentials $V$ and $V'$ are related to each other as
\begin{eqnarray} \label{new_GF2}
  \!\!\!\! G'=A^{-1} G,\;\;  A= I+(I-G)(\textrm{exp}(V'-V)-I),
\end{eqnarray}
where $I$ denotes the unit matrix.
However, in contrast to the HF-QMC, Eq.~(\ref{new_GF2}) now contains off-diagonal in the spin indices elements.
In the case of a single auxiliary spin-flip, the fast matrix update algorithm has the same form as in the HF-QMC method.
Note, however, that all equations have the matrix form in the spin indices.
The ratio of stochastic weights for two configurations which differ by the quantization axis direction
can be calculated as
\begin{eqnarray} \label{rot_ratio}
  \dfrac{\textrm{det}[G(s,\theta,\phi)]}{\textrm{det}[G'(s,\theta',\phi')]} = \textrm{det}[A].
\end{eqnarray}
If the new quantization axis direction defined by angles $\theta'$ and $\phi'$ is accepted,
the corresponding Green's function ${G'(s,\theta',\phi')}$ is calculated via the full update procedure
using Eq.~(\ref{new_GF2}).

The physical meaning of the proposed rotationally invariant algorithm
can be expressed as averaging over all possible directions of fluctuating
spin polarization in the 3$d$ shell, in contrast to polarization along the 
$z$-axis only in the density-density Hamiltonian. 
Thereby this technique allows one to take into account not only the longitudinal spin
fluctuations but also the transverse ones.

\section{Results and discussion\label{sec:results}}

In this section, we first present results of our model 
calculations in comparison with previous studies. In particular, we employ 
the proposed approach to compute the two- and three-band Hubbard models on the 
infinite-dimensional Bethe lattice. We benchmark these calculations with the 
previously published results of the CT-QMC computations.~\cite{Antipov_3band_model}
To study the role of symmetry of the exchange interaction we compare our 
results with those obtained by employing the density-density form
of the local Coulomb interaction.
In particular, we compute the
uniform magnetic susceptibility $ \chi (T) = dM(T)/ dH_z$ by
calculating magnetization $M (T) = \sum_m (n_{m{\uparrow}} - n_{m{\downarrow}})$
induced by the external magnetic field $H_z$ applied along the $z$ axis
(these calculations include the polarization of the impurity Weiss field).
In our calculations, we adopted a few magnetic fields in the range from 0.01 
to 0.04 eV to ensure linearity of response.
The temperature dependence of magnetic susceptibility $\chi (T)$ is fitted to 
the Curie-Weiss law $ \chi (T)=C/(T-T_\textrm{C})$, where $C$ is a material-specific constant 
and $T_\textrm{C}$ is the Curie temperature.
Next we investigate the electronic and magnetic properties
of paramagnetic iron and nickel. We outline the importance of the correct spin 
rotational symmetry of the exchange interaction to describe the
properties of these materials. These results are presented in Secs. III B and III C.

\subsection{Two- and three-band models}

In recent years the properties of the two- and three-orbital Hubbard 
models have been extensively investigated by using dynamical mean-field 
approach.~\cite{Prushke_2band_model,Sakai,Bethe_studies,Antipov_3band_model}
Here we would like to refer to Ref.~\onlinecite{Antipov_3band_model} where
the two- and three-orbital Hubbard models on the infinite-dimensional Bethe lattice
were studied by means of the continuous-time quantum Monte Carlo method
with full rotationally invariant Coulomb interaction.
In particular, it was established that the effect of spin-flip interactions 
considerably depends on a band filling. It was shown that the Mott-Hubbard physics 
dominates due to the strong effective Coulomb interaction $U$ in the particle-hole 
symmetric case, while away from half-filling formation of local magnetic moments is 
more plausible due to the Hund's exchange. In agreement with previous 
studies,\cite{Sakai,Antipov_2band_model,jqmc} it 
was also shown that using the density-density form of the Coulomb interaction 
results in an overestimation of the magnetic transition temperature.
For the benchmark purposes we here reproduce these calculations by employing the proposed 
rotationally invariant \hfqmc method.

\begin{figure}[t]
\centering
\includegraphics[clip=true, width=0.46\textwidth]{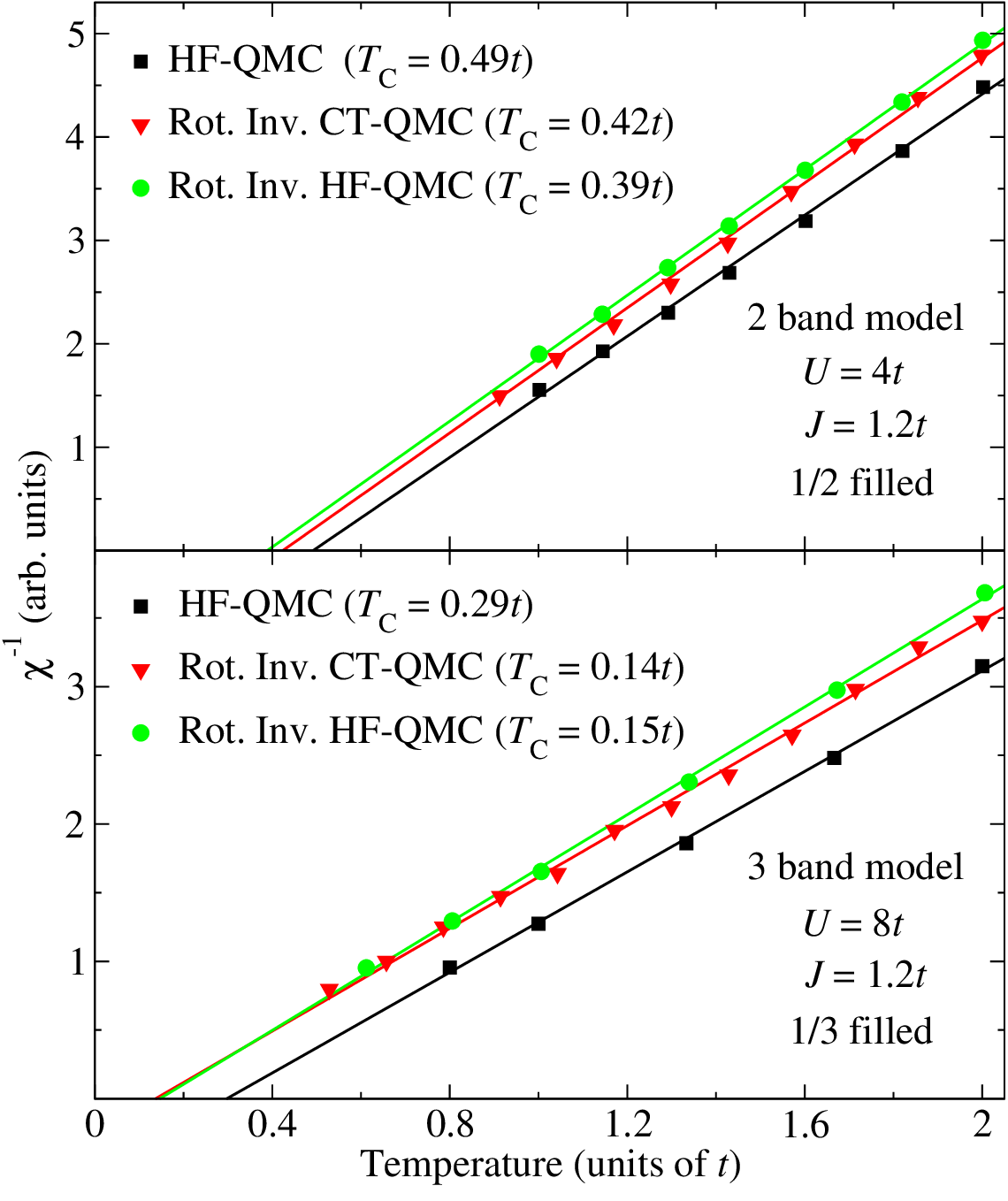}
\caption{(Color online)
Temperature dependence of the inverse uniform magnetic susceptibility as obtained by DMFT
for the two-band (upper panel) and three-band (lower panel) Hubbard models on the 
infinite-dimensional Bethe lattice. The straight lines depict the least-squares fit to 
the Curie-Weiss law. The extracted Curie temperatures and effective local magnetic moments 
are presented in Table~\ref{tab:model}. The CT-QMC results were taken from Ref.~\onlinecite{Antipov_3band_model}.
\label{fig:model_susc}}
\end{figure}

In Fig.~\ref{fig:model_susc} (upper panel) we present our results for the inverse 
uniform magnetic susceptibility calculated for the two-band Hubbard model on the 
infinite-dimensional Bethe lattice. We have chosen the same set of parameters 
as in Ref.~\onlinecite{Antipov_3band_model}. Namely, we consider the two-band
model at half-filling with ${U=4t}$ for Coulomb interaction 
and ${J=1.2t}$ for Hund's exchange, where $t$ is a half of the non-interacting 
bandwidth. The calculated Curie temperatures are \,0.49$t$ and 0.39$t$ for the conventional 
and rotationally invariant HF-QMC, respectively. These findings are in good
quantitative agreement with the results of the CT-QMC calculations
which give ${T_\textrm{C} \sim 0.49t}$ and 
$0.42t$ for the density-density and rotationally invariant interactions.
Similar calculations with ${U=8t}$ (see Table~\ref{tab:model}) give ${T_\textrm{C} \sim 0.49t}$ and $0.36t$
for the density-density and rotationally invariant HF-QMC methods, respectively.
We notice that the effective local moments calculated by the conventional and rotationally 
invariant HF-QMC methods agree well with those obtained by the CT-QMC.

To proceed further, we investigate the properties of the three-band Hubbard model with 
${U=8t}$ and ${J=1.2t}$ at one-third electron filling (two-electron occupancy). 
In Fig.~\ref{fig:model_susc} (lower panel) we present our results for the inverse uniform 
magnetic susceptibility. Our results for $U=4t$ and half-filling are summarized 
in Table~\ref{tab:model}.
For both sets of parameters, the broken rotational symmetry of the Coulomb interaction
leads to an overestimation of the Curie temperature.
However, in agreement with Ref.~\onlinecite{Antipov_3band_model}, this overestimation is found 
to be more pronounced at the $1/3$ filling than at the half-filling. 
This is due to the Hund's exchange interaction which plays a dominating role away from half-filling.
The Curie temperatures calculated by the rotationally invariant HF-QMC algorithm are
$0.69t$ and $0.15t$ for the $1/2$ and $1/3$ electron filling, respectively.
These findings are in good quantitative agreement with the results of rotationally invariant 
CT-QMC calculations which give $0.70t$ and $0.14t$ for the half-filling and one-third filling, 
respectively.
Our findings clearly indicate that the Curie temperature is overestimated by 
employing the density-density form of Coulomb interaction. Hence, the retaining 
of spin rotational symmetry is crucial for the correct description of the magnetic transition 
temperature.

We find good quantitative agreement between the results obtained by rotationally invariant 
HF-QMC and CT-QMC methods. This demonstrates the validity of our method for accurate description 
of the magnetic properties of correlated electron systems. We note that transverse spin fluctuations 
can be regarded as an important source of magnetic response softening. It is expected that the 
proper treatment of the spin rotational symmetry of Coulomb interaction is even more important in 
the five-band case.

\begin{table}[t]
  \caption{Curie temperatures (in units of~$t$) and effective local magnetic moments (in~$\mu_B$)
  as obtained by DMFT for the two- and three-band Hubbard models on the infinite-dimensional Bethe lattice.
  The results corresponding to the density-density interaction are denoted as HF-QMC and \mbox{CT-QMC}.
  The calculations were carried out with ${J = 1.2t}$. The \mbox{CT-QMC} results were taken 
  from Ref.~\onlinecite{Antipov_3band_model}.
  \label{tab:model}}
  \begin{ruledtabular}
    \begin{tabular}{cccccc}
      Bands  & Filling  &  U/t &  Impurity solver  & $T_\textrm{C}$ & $\mu_\textrm{eff}$  \\
\hline
        2    &   1/2    &   4  &      CT-QMC       &      0.49      &   2.02     \\
             &          &      &      HF-QMC       &      0.49      &   2.02     \\
             &          &      &  Rot. Inv. CT-QMC &      0.42      &   1.99     \\
             &          &      &  Rot. Inv. HF-QMC &      0.39      &   1.98     \\
\hline
        2    &   1/2    &   8  &      CT-QMC       &      0.50      &   2.21     \\
             &          &      &      HF-QMC       &      0.49      &   2.22     \\
             &          &      &  Rot. Inv. CT-QMC &      0.40      &   2.20     \\
             &          &      &  Rot. Inv. HF-QMC &      0.36      &   2.19     \\
\hline
        3    &   1/2    &   4  &      CT-QMC       &      0.83      &   2.41     \\
             &          &      &      HF-QMC       &      0.84      &   2.42     \\
             &          &      &  Rot. Inv. CT-QMC &      0.70      &   2.41     \\
             &          &      &  Rot. Inv. HF-QMC &      0.69      &   2.35     \\
\hline
        3    &   1/3    &   8  &      CT-QMC       &      0.27      &   2.54     \\
             &          &      &      HF-QMC       &      0.29      &   2.56     \\
             &          &      &  Rot. Inv. CT-QMC &      0.14      &   2.55     \\
             &          &      &  Rot. Inv. HF-QMC &      0.15      &   2.48     \\
    \end{tabular}
 \end{ruledtabular}
\end{table}

\subsection{$\alpha$ iron}

Elemental iron is one of the oldest and experimentally best studied itinerant ferromagnets.
Various properties of iron can be understood on the basis of band-structure calculations.~\cite{Fe_LDA,Corso} 
In particular, these calculations provide a good description of the low-temperature ferromagnetic phase of Fe.
However, applications of conventional band-structure techniques to describe the 
properties of paramagnetic iron, in particular, close to the $\alpha-\gamma$ phase transition,
do not lead to satisfactory results. This is mainly due to the presence of local magnetic moments 
above the Curie temperature which are important for quantitative description of paramagnetic state.
In this respect, the LDA+DMFT approach provides the best formalism which allows one to unify the localized 
and itinerant electron behavior in metallic magnets.

\begin{figure}[b]
\centering
\includegraphics[clip=true, width=0.46\textwidth]{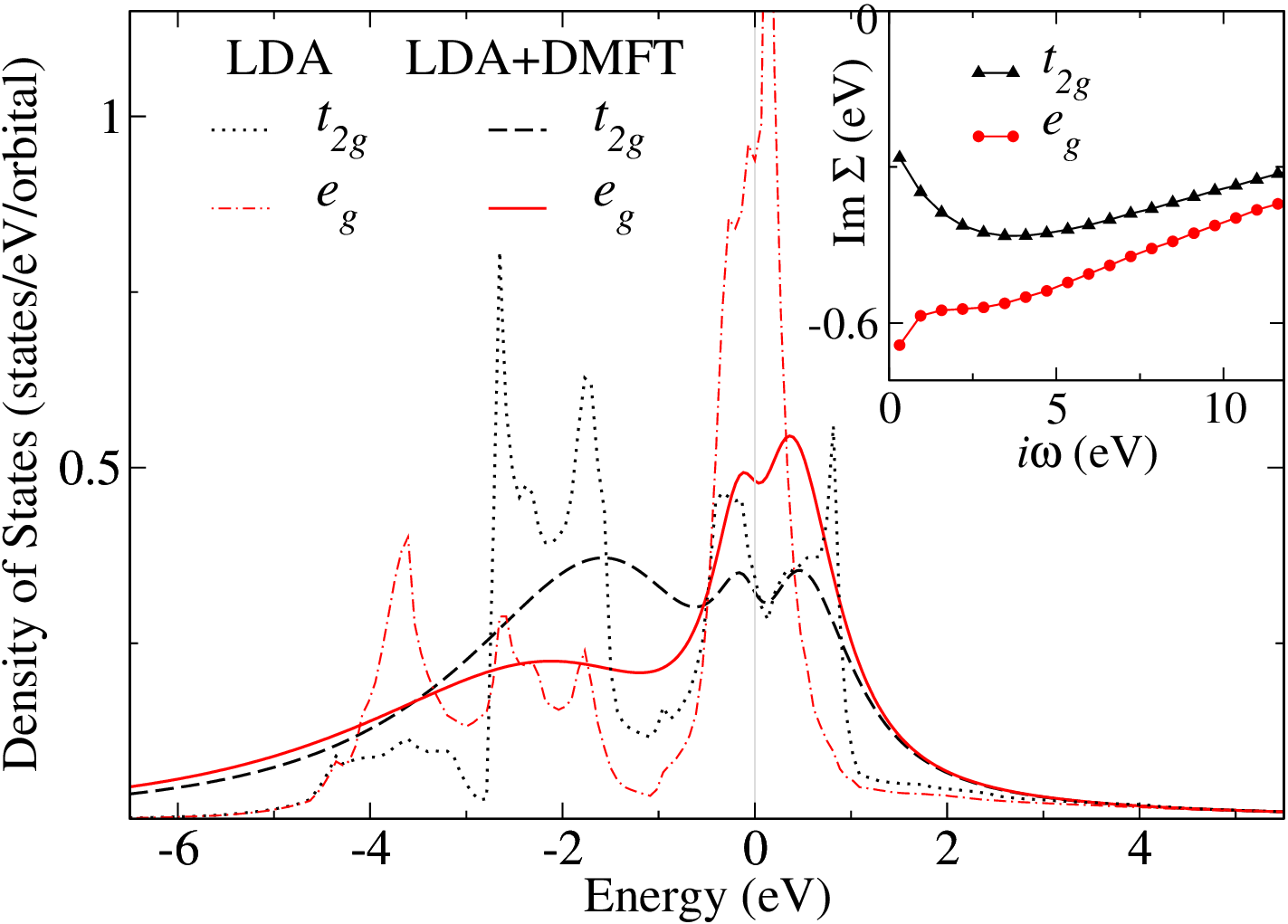}
\caption{(Color online) 
Partial $t_{2g}$ (black) and $e_g$ (red) densities of states for paramagnetic $\alpha$ iron
as obtained by the rotationally invariant HF-QMC within LDA+DMFT in comparison with the 
non-magnetic LDA results. The Fermi level is indicated by the vertical (gray) line at zero 
energy. Inset: imaginary parts of the obtained self-energies.
\label{fig:fe_dos_1spin}}
\end{figure}

Applications of LDA+DMFT have shown to provide a good quantitative description of the electronic, 
magnetic, and structural properties
of iron.~\cite{Katsnelson99,Katsnelson00,Licht01,Katanin10,Fe_Leonov,Benea,Chioncel_Fe_Ni,Grechnev}
However, an agreement was achieved only in terms of the reduced temperature $T/T_C$, while the calculated 
Curie temperature $T_C$ was found to be almost twice larger than the experimental value of $1043$~K 
(Ref.~\onlinecite{susc_exp}).
Recently the properties of iron have been investigated by means of $\vec{J}$-QMC method,
which uses the static approximation for the charge degrees of freedom
and treats the exchange Coulomb interaction in the rotationally invariant form.~\cite{jqmc} 
These calculations indicate that substantial part of the Curie temperature overestimation comes 
from the approximate (density-density) treatment of the exchange Coulomb interaction.

We now calculate the electronic structure and magnetic properties of paramagnetic 
bcc iron by employing the LDA+DMFT implemented with the rotationally invariant HF-QMC method.
We first calculate the non-magnetic LDA electronic structure of $\alpha$ iron using
the tight-binding linear muffin-tin orbital (TB-LMTO) approach.~\cite{LMTO}
For these calculations we adopt lattice constant ${a= 2.866}$~\AA. We construct an 
effective low-energy Hamiltonian in the basis of Fe $spd$ Wannier orbitals using the
$N$th-order muffin-tin ($N$MTO) method.~\cite{Andersen00}
Here we adopt common definitions for the screened Coulomb interaction
and Hund's exchange parameters in the $3d$~shell, namely,
${U \equiv F^0}$ and ${J \equiv (F^2+F^4)/14}$, where $F^0$, $F^2$, and $F^4$ are the Slater integrals. 
We take ${U=2.3}$~eV and ${J=0.9}$~eV in accordance with the previous estimations~\cite{Katsnelson99,Licht01,Coco05,Katanin10}
and solve the five-orbital impurity problem within DMFT.
\begin{figure}
\centering
\includegraphics[clip=true, width=0.46\textwidth]{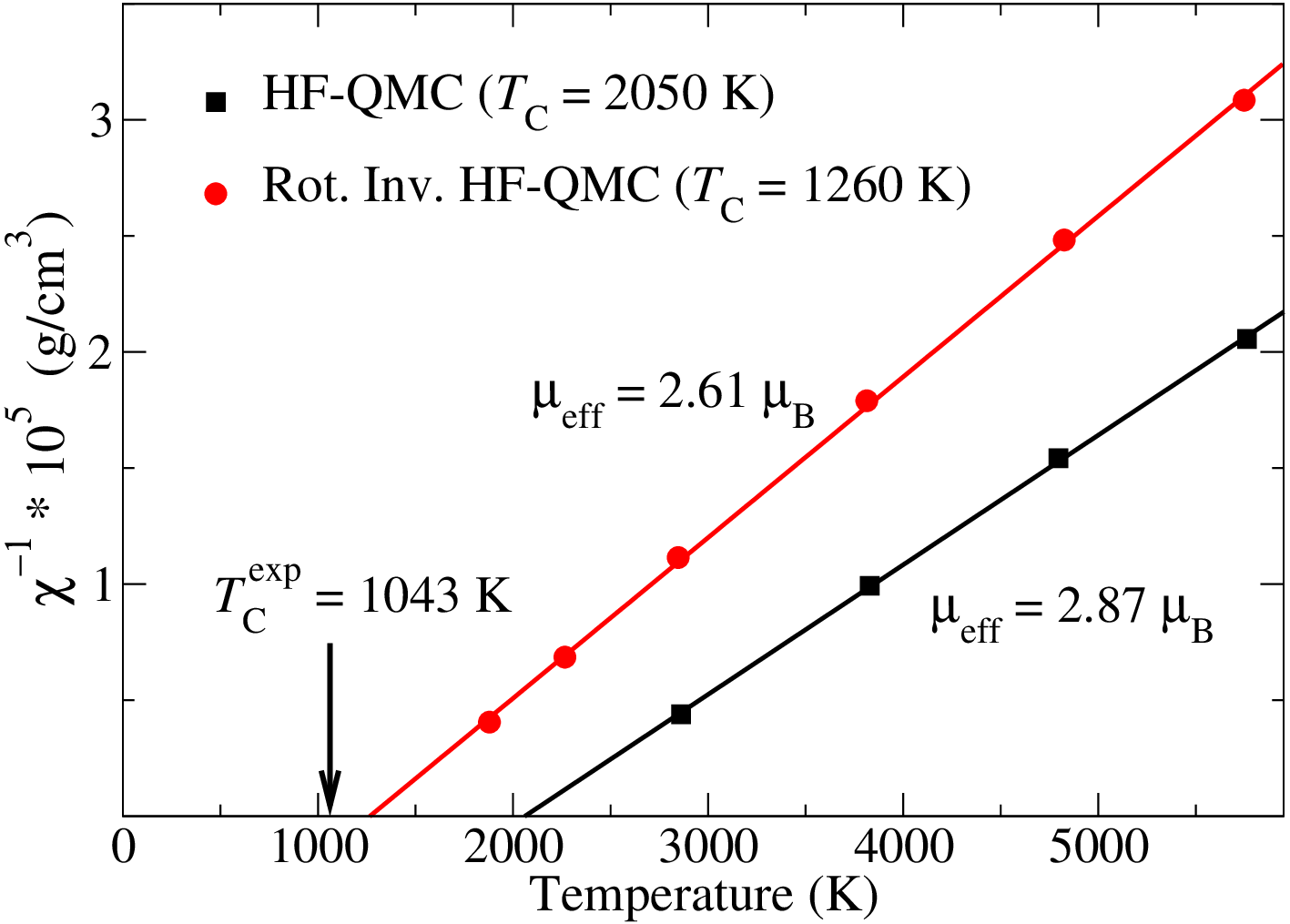}
\caption{(Color online) 
Temperature dependence of the inverse uniform magnetic susceptibility for $\alpha$ iron 
as obtained by the LDA+DMFT. The straight lines depict the least-squares fit to the Curie-Weiss law.
The experimental value $T_\textrm{C}^\textrm{exp} = 1043$ K is denoted by the (black) arrow.
The experimental value of the effective local magnetic moment 
is ${\mu_\textrm{eff}^\textrm{exp} = 3.13 \mu_B}$ (Ref.~\onlinecite{susc_exp}).
\label{fig:fe_susc}}
\end{figure}

In Fig.~\ref{fig:fe_dos_1spin} we present the partial density of states
and the corresponding imaginary parts of the self-energies obtained by the 
rotationally invariant HF-QMC method at $T = 1160$ K in comparison with 
the non-magnetic LDA results. 
Our calculations reproduce the splitting in the density of states of the 
$e_g$ orbitals near the Fermi level, which is absent in the LDA calculations.
This result agrees well with the previous calculations~\cite{Katsnelson99,Katanin10,jqmc}
as well as with the experimental data and is one of the characteristic features of $\alpha$ 
iron. The calculated self-energy for the $t_{2g}$ states exhibits a Fermi-liquid-like behavior, 
whereas the $e_g$ self-energy diverges at low frequencies. Our calculations with ${J=0}$ recover 
the Fermi-liquid-like behavior for the $e_g$ states resulting in the suppression of the splitting.
This indicates that the splitting can be attributed to the exchange Coulomb interaction.~\cite{Katanin10}
We found no evidence for the Hubbard subbands formation in the calculated quasiparticle spectrum.
Thereby we conclude that in $\alpha$ iron the correlation effects are mainly affected by
the strength of the Hund's coupling $J$ rather than by the Coulomb interaction $U$ which 
allow us to refer $\alpha$ iron as a Hund's metal. We note that importance of the Hund's 
exchange in multiorbital systems has been recently studied in Ref.~\onlinecite{Medici}.

The temperature dependence of the inverse uniform magnetic susceptibility
calculated by the LDA+DMFT shows a linear behavior at high temperatures
(Fig.~\ref{fig:fe_susc}) in accordance with the Curie-Weiss law. It is clearly 
seen that the \hfqmc limited to the Ising-type exchange interaction substantially 
overestimates the Curie temperature value and yields ${T_\textrm{C} \sim 2050}$~K. 
The rotationally invariant \hfqmc method gives ${T_\textrm{C} \sim 1260}$~K, which 
is in good agreement with the experimental value of $1043$~K (Ref.~\onlinecite{susc_exp}).
The calculated values of the effective local moment extracted from the uniform magnetic susceptibility
are ${\mu_\textrm{eff} \sim 2.87~\mu_\textrm{B}}$ and $2.61~\mu_\textrm{B}$ for the
HF-QMC with density-density and rotationally invariant interactions, respectively.
Our estimates are in reasonable agreement with the experimental value of $3.13~\mu_\textrm{B}$ 
(Ref.~\onlinecite{susc_exp}), which appears to be underestimated by both methods.

We note that preserving the spin rotational symmetry turns out to be crucial 
for quantitative description of magnetic properties of correlated materials.
In particular, for $\alpha$ iron this leads to substantial improvement of 
the magnetic transition temperature value.

\subsection{Nickel}

Elemental nickel is another example of itinerant electron ferromagnets which 
together with iron serves as a benchmark material for electronic structure methods.
Various low temperature properties of nickel can be understood by employing 
standard band-structure approaches.~\cite{Corso,Ni_LDA}
Nevertheless, these techniques generally
fail to reproduce many characteristic features of nickel, such as an existence of satellite 
structure~\cite{Ni_satellite} at $-6$ eV, $3d$ electron bandwidth,~\cite{Ni_bandwidth_and_exchange}
and the value of exchange splitting.~\cite{Ni_bandwidth_and_exchange} 
Applications of LDA+DMFT to study the electronic and magnetic properties of nickel 
have given a good quantitative description of many of these 
phenomena.~\cite{ni_studies,Katsnelson99,Licht01,Katsnelson02,Benea,Chioncel_Fe_Ni,Grechnev}
The calculated magnetic 
properties of nickel are shown to be in good agreement with experiment. In contrast to iron, 
the overestimation of the magnetic transition temperature for nickel by LDA+DMFT with the 
Ising-type exchange interaction is not so significant.~\cite{Licht01,Katsnelson04} However, as 
demonstrated below, preserving the spin rotational symmetry leads to the underestimation of
the Curie temperature.

We now compute the electronic structure and magnetic properties of paramagnetic 
nickel by employing the LDA+DMFT implemented with rotationally invariant HF-QMC method.
To obtain the non-magnetic LDA electronic structure of fcc nickel we employ the tight-binding 
linear muffin-tin orbital approach.
The calculations were performed for the 
lattice constant $a= 3.524$~\AA. Using these results we construct an effective low-energy 
Hamiltonian in the basis of Ni $spd$ Wannier orbitals by employing the $N$MTO method. 
In accordance with the previous estimations,~\cite{Katsnelson02,Benea} we take ${U=2.3}$~eV and ${J=1.0}$~eV for the 
screened Coulomb interaction and Hund's exchange, respectively.
\begin{figure}
\centering
\includegraphics[clip=true, width=0.46\textwidth]{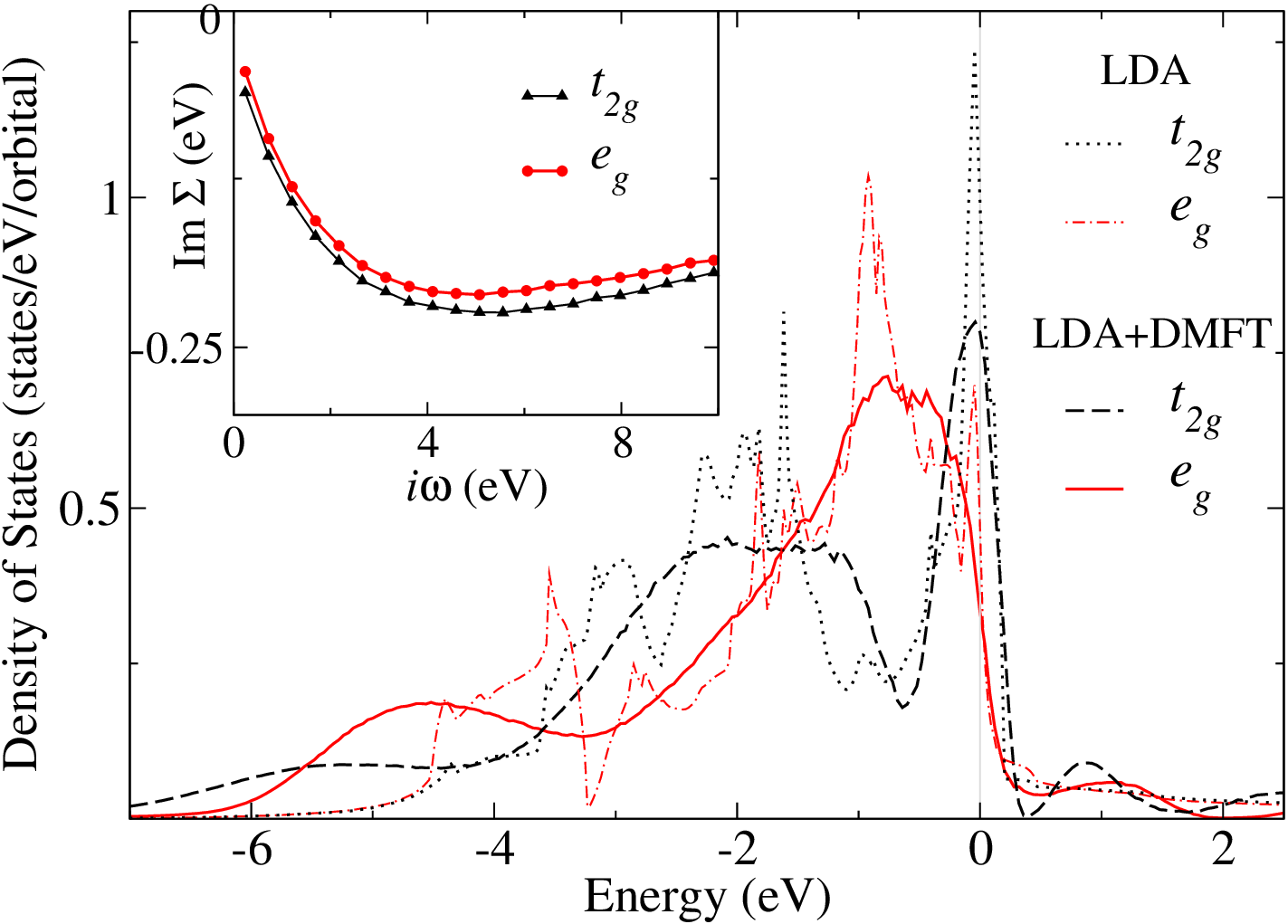}
\caption{(Color online) 
Partial densities of states for paramagnetic Ni obtained by rotationally invariant 
HF-QMC within LDA+DMFT in comparison with the LDA ones. The Fermi level is indicated 
by the vertical (gray) line at zero energy. Inset: imaginary parts of the obtained self-energies
\label{fig:ni_dos_1spin}}
\end{figure}

\begin{figure}[t]
\centering
\includegraphics[clip=true, width=0.46\textwidth]{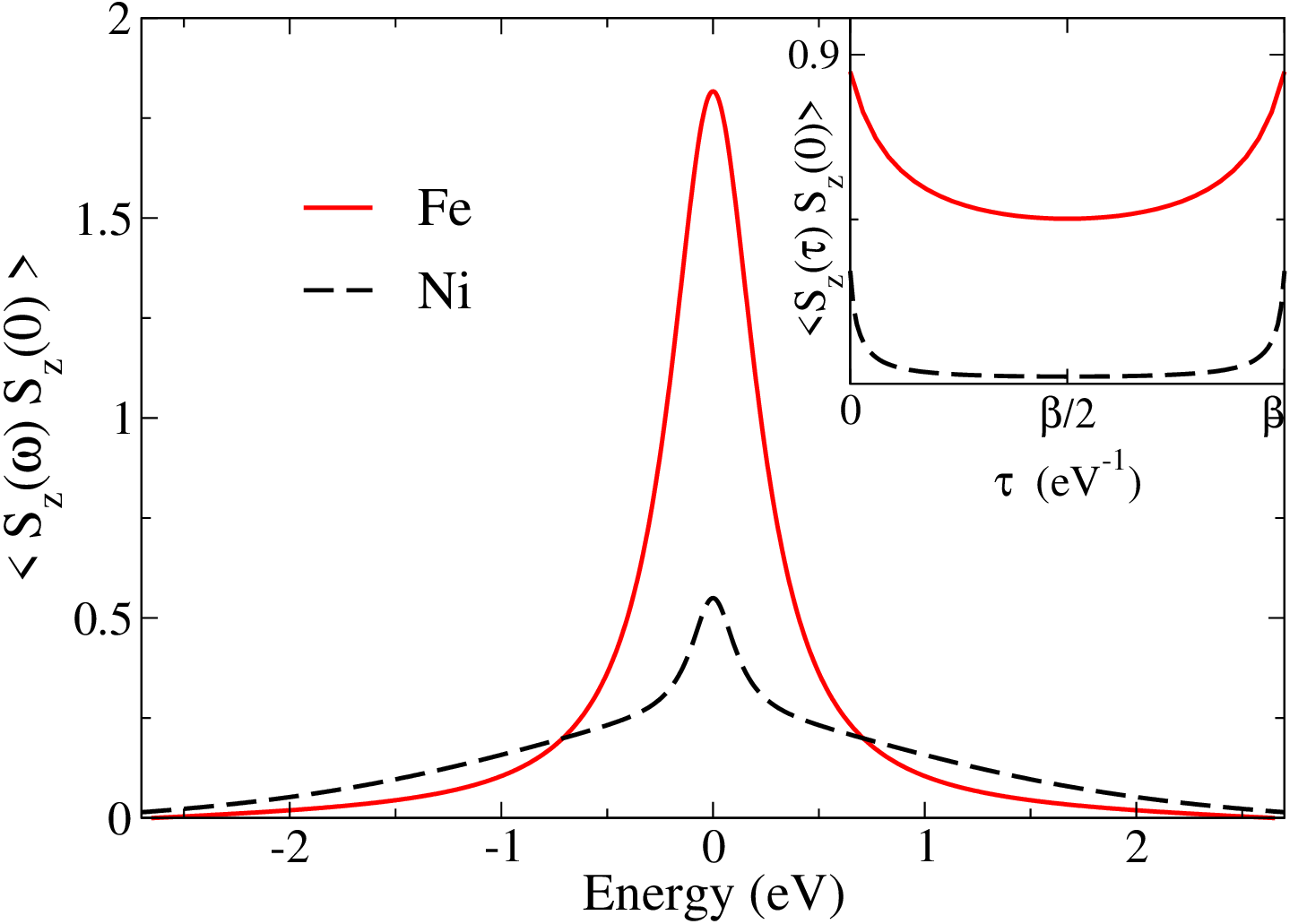}
\caption{(Color online)
Spin-spin correlation functions on the real and imaginary (inset) energy axes for 
$\alpha$ iron and nickel calculated by LDA+DMFT at ${T=2.5\, T_\textrm{C}}$.
\label{fig:sisj}}
\end{figure}

In Fig.~\ref{fig:ni_dos_1spin} we present the partial densities of states
and the imaginary parts of the self-energies obtained by the rotationally invariant 
algorithm at ${T = 1160}$ K in comparison with the non-magnetic LDA results.
The inclusion of electronic correlations results in a small reduction of the bandwidth 
of nickel with respect to the non-magnetic LDA result. In addition, a satellite-like 
structure emerges at about $-5.5$~eV. In contrast to iron, the obtained self-energies 
for both the $t_{2g}$ and $e_g$ orbitals exhibit the Fermi-liquid-like behavior.
This can also be seen from the calculated amplitudes of the effective damping ${\Im\Sigma(0)}$ which 
are $-0.03$~eV and $-0.02$~eV for the $t_{2g}$ and $e_g$ states of nickel, respectively, while
${\Im\Sigma(0)\sim -0.24}$~eV for the $t_{2g}$ states of $\alpha$ iron. This indicates a more coherent 
nature of the electronic properties of nickel in comparison with iron. To quantify this qualitative 
difference \cite{Licht01,Katsnelson04} we calculate spin-spin correlation functions for iron and nickel.
In Fig.~\ref{fig:sisj} we present the impurity spin-spin correlation functions 
on the real and imaginary energy axes calculated for $\alpha$ iron and nickel at ${T = 2.5\, T_\textrm{C}}$
($T_\textrm{C}$ refers here to the corresponding calculated value of the Curie temperature).
The height of peak on the real energy axis can be interpreted as a value of the local magnetic moment. 
The pronounced peak for iron indicates the presence of well localized magnetic moments
above $T_\textrm{C}$, while magnetism of nickel is more itinerant.
 
\begin{figure}
\centering
\includegraphics[clip=true, width=0.46\textwidth]{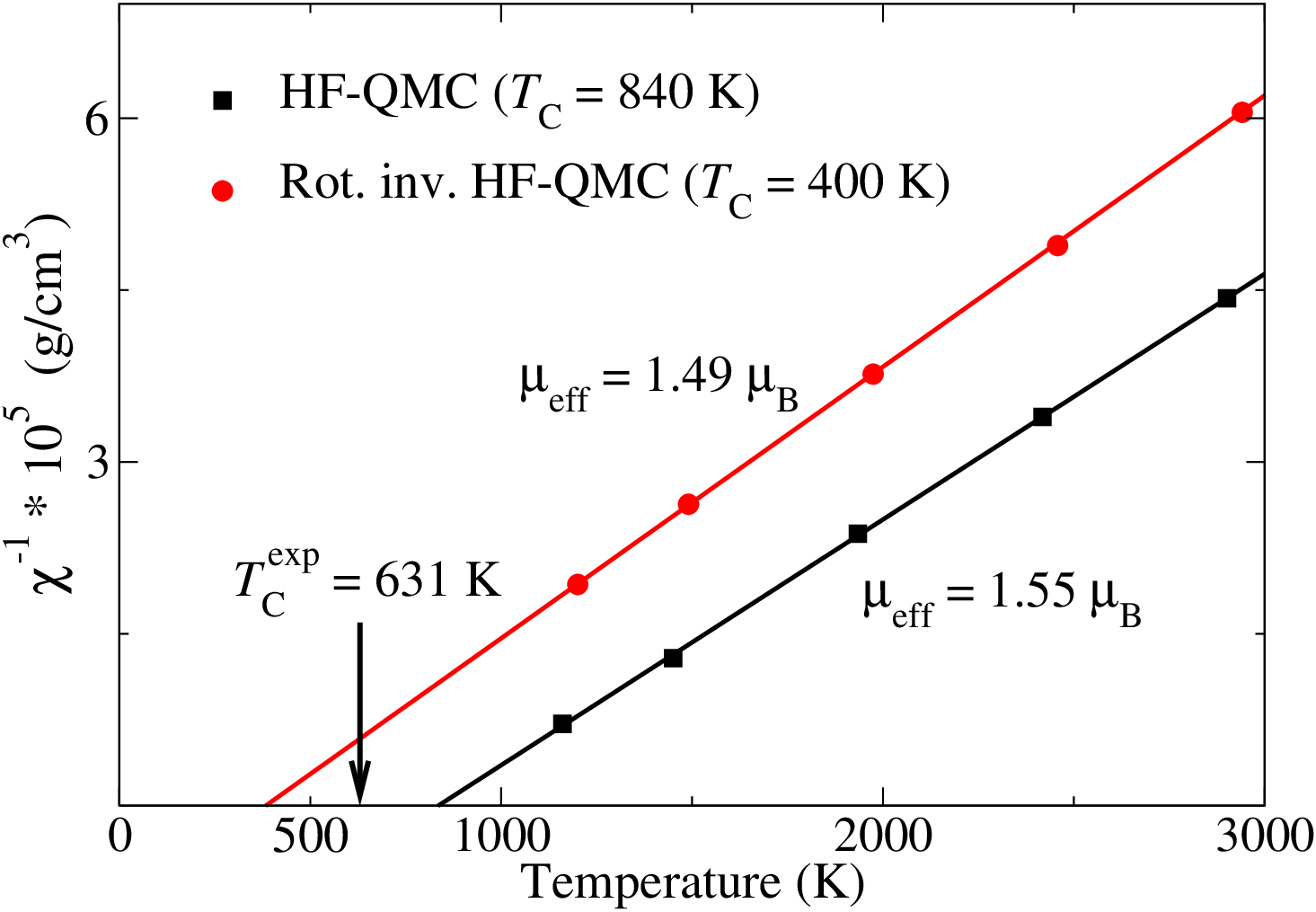}
\caption{(Color online) 
Temperature dependence of the inverse uniform magnetic susceptibility for Ni as obtained by LDA+DMFT.
The straight lines depict the least-squares fit to the Curie-Weiss law. The experimental value 
$T_\textrm{C}^\textrm{exp}=631$~K is denoted by the (black) arrow. The experimental value of the 
effective local magnetic moment is $\mu_\textrm{eff}^\textrm{exp} = 1.62~\mu_B$ (Ref.~\onlinecite{susc_exp}).
\label{fig:ni_susc}}
\end{figure}

In Fig.~\ref{fig:ni_susc} we show the temperature dependence of the inverse uniform magnetic 
susceptibility calculated by LDA+DMFT. From these data we estimate the values of the Curie temperature.
By employing the density-density approximation we find ${T_\textrm{C} \sim 840}$ K, whereas
the inclusion of the spin rotational symmetry leads to an almost twice smaller value of about 400~K.
Both results are in reasonable agreement with the experimental value of 631~K. To proceed further, 
we compute the effective local magnetic moments by the HF-QMC and rotationally invariant 
methods which give ${\mu_\textrm{eff} \sim 1.55~\mu_\textrm{B}}$ and ${1.49~\mu_\textrm{B}}$, respectively.
These findings are in good agreement with the experimental value of $1.62~\mu_B$~(Ref.~\onlinecite{susc_exp}).
The fact that in $\alpha$ iron the LDA+DMFT with proper spin rotational symmetry results in 
an overestimated value of $T_C$, while it appears to be underestimated in nickel, can be dealt 
with non-local effects which are neglected in DMFT or with the more itinerant nature of magnetism 
in nickel than in $\alpha$ iron.

\section{Conclusions\label{sec:conclusions}}

In conclusion, we have presented an implementation of the LDA+DMFT
approach which allows one to take into account the spin rotational symmetry 
of the exchange Coulomb interaction. The computational scheme is based on 
extension of the Hirsch-Fye quantum Monte Carlo algorithm in which the spin 
rotational invariance of Hund's exchange is approximated by averaging over all 
possible directions of the spin quantization axis. The proposed approach provides 
a robust and computationally efficient method which allows us to compute high 
temperature electronic properties of the five-orbital systems. We have used 
this approach to perform benchmark calculations for the two- and three-band Hubbard models
on the infinite-dimensional Bethe lattice. Our results agree quantitatively well 
with those obtained using the continuous-time quantum Monte Carlo technique.
The proposed method is employed to compute the electronic and magnetic properties 
of paramagnetic $\alpha$ iron and nickel. The obtained Curie temperatures agree 
well with experiment. Our results indicate that the density-density approximation 
for the Coulomb interaction leads to a substantial overestimation of the 
magnetic transition temperature.

\begin{acknowledgments}
 
The authors thank T.~Costi, E.~Dagotto, A.~Liebsch, A.~Poteryaev, and S.~Sakai for valuable comments.
The authors are also grateful to A.~Antipov for providing the CT-QMC data.
This work was supported by the Russian 
Foundation for Basic Research (Projects Nos. 10-02-00046a, 12-02-91371-CT$\_$a, 12-02-31207), 
the Ministry of education and science of Russian Federation through projects 14.A18.21.0076, 12.740.11.0026, the fund of the 
President of the Russian Federation for the support of scientific schools 
NSH-6172.2012.2, the Dynasty Foundation,
Program of the Russian Academy of Science Presidium “Quantum microphysics of condensed matter” 12-P-2-1017, 12-M-23-2020.
Support by the Deutsche Forschergemeinschaft through TRR~80 and FOR~1346 
is gratefully acknowledged.

\end{acknowledgments}


\begin{thebibliography}{99}
\bibitem{Footnote} 
Strictly speaking, this parametrization of the Hamiltonian is appropriate for the two- and three-orbital systems.
The five-band calculations require three effective Slater integrals ($F^0$, $F^2$, and $F^4$)
which can be linked to the $U$ and $J$ parameters.
\end{thebibliography}
\end{document}